\documentclass[11pt]{article}
\usepackage{amsfonts}
\usepackage{amsmath}
\usepackage{latexsym}
\usepackage{times}

\usepackage{epsfig}
\usepackage{multirow}
\usepackage{hyperref}
\usepackage{url}
\newcommand{\ket}[1]{|\, #1\rangle}
\newcommand{\bra}[1]{\langle #1\,|}

\newcommand\beq{\begin{equation}}
\newcommand\eeq{\end{equation}}
\newcommand\bea{\begin{eqnarray}}
\newcommand\eea{\end{eqnarray}}

 \def\squarebox#1{\hbox
to #1{\hfill\vbox to #1{\vfill}}}
\def\qed{\hspace*{\fill}\vbox{\hrule\hbox{\vrule\squarebox{.667em}\vrule}\hrule}
} 

\newcommand{\ba}{\begin{array}}
\newcommand{\ea}{\end{array}}

\newtheorem{defi}{Definition}

\def\tr{\mbox{\sf tr}}

\begin{document}

\title{Security Trade-offs in Ancilla-Free Quantum Bit Commitment in the Presence of Superselection Rules}
\author{David P. DiVincenzo, John A. Smolin and Barbara M. Terhal \footnote{IBM Watson Research Center,
P.O. Box 218, Yorktown Heights, NY 10598, USA.}}

\date{\today}

\maketitle

\begin{abstract}
Security trade-offs have previously been established for one-way
bit commitment. We study this trade-off in two superselection
settings. We show that for an `abelian' superselection rule
(exemplified by particle conservation) the standard trade-off
between sealing and binding properties still holds. For the
non-abelian case (exemplified by angular momentum conservation)
the security trade-off can be more subtle, which we illustrate by
showing that if the bit commitment is forced to be ancilla-free,
an asymptotically secure quantum bit commitment is possible.
\end{abstract}

{PACS}: 03.670.-a, 03.67.Dd \\

\section{Introduction}

The question of whether the no-go result for quantum bit
commitment \cite{mayers:qbcsecurity}, \cite{LC:qbcsecurity},
\cite{LC:qbcsecurity2} remains valid in the presence of
superselection rules has been addressed by several recent papers
\cite{VC:ss}, \cite{KMP:ss}, \cite{mayers:ss}. The most general
result is described in Ref. \cite{KMP:ss}, where the authors prove
that secure quantum bit commitment is impossible even with general
superselection rules. The authors of this paper distinguish
security in the case of `abelian' superselection rules (such as
particle conservation) and `nonabelian' superselection rules in
which the imposed symmetry is described by a non-abelian group. In
both cases the authors prove the impossibility of establishing a
secure bit-commitment protocol. In this paper we point out that
the no-go result for the non-abelian case is fairly non-trivial;
we show that if the cheating strategies are forced to be
ancilla-free, an asymptotically secure protocol can be found. In
contrast, for the abelian case (or in the absence of
superselection rules) the `no-ancilla' enforcement does not alter
the security trade-offs for a one-way bit commitment (called a
`purification bit commitment' in Ref. \cite{SR:bc}).


\section{One-way Bit Commitment Protocol}

We repeat the definition in Ref. \cite{SR:bc} of this important class
of bit commitment (BC) protocols:

\begin{defi}[One-Way Bit Commitment Protocol, BC] \cite{SR:bc}
In this protocol, Bob begins with no quantum state of his own,
Alice begins with a two-part Hilbert space ${\cal H}_p\otimes{\cal
H}_t$ (``proof" and ``token"). Alice chooses to commit to bit $b$.
Alice prepares one of two orthogonal states $\ket{\chi_b}$ in her
total Hilbert space. In the {\rm commit} phase, Alice transmits
to Bob the state in ${\cal H}_t$ which we denote by $\rho_b$;
in the {\rm unveiling} phase, Alice transmits the state in ${\cal H}_p$ to Bob; Bob determines the
committed bit by projectively measuring the state using orthogonal
projectors $\{\Pi_0,\Pi_1,\Pi_{fail}\}$.
\end{defi}

This is certainly not the most general quantum bit commitment
protocol, which would permit more than one round of communication.
The security trade-offs of one-way BC protocols have been
described as follows, see Ref. \cite{SR:bc}. Two scenarios are
considered: 1) Alice is honest, but Bob tries to cheat by learning
the bit in the commit phase. 2) Bob is honest, but Alice tries to
cheat by changing her committed bit after the commit phase.  In
case 1), Bob is trying to make his ``information gain" $G(S^B)$
nonzero, defined as the difference between his probability of
estimating Alice's commitment correctly in the commit phase when
he employs cheating strategy $S^B$, and when he is honest,
\begin{equation}
G(S^B)=P_E(S^B)-1/2.
\end{equation}
In case 2), Alice is trying to make her ``control" $C(S^A)$
nonzero, defined as the difference between her probability of
unveiling whatever bit she desires when she implements $S^A$, and
when she is honest,
\begin{equation}
C(S^A)=P_U(S^A)-1/2.
\end{equation}
Then, the security of any given protocol can be characterized by
the maximum of these two quantities:
\begin{eqnarray}
G^{\mbox{max}}&\equiv&\max_{S^B}G(S^B),\nonumber\\
C^{\mbox{max}}&\equiv&\max_{S^A}C(S^A).
\end{eqnarray}
For one-way bit commitment protocols it has been established
\cite{SR:bc} that \bea
G^{\mbox{max}}=\frac{1}{2} D(\rho_0,\rho_1)={1\over 4}\tr\,|\rho_0-\rho_1|. \label{dist} \\
C^{\mbox{max}}=\frac{1}{2} F(\rho_0,\rho_1)={1\over
2}\tr\,|\sqrt{\rho_0}\sqrt{\rho_1}| \label{fid}, \eea where
$F(.,.)$ is the fidelity function and $D(.,.)$ is the
trace-distance function. Notably, the cheating strategies that
achieve these optima do not make use of ancillas: Alice's cheating
strategy is the creation of a $b-$independent state potentially
followed by unitary rotation, whereas Bob's cheating strategy is a
complete von Neumann measurement projecting in the eigenbasis of
$\rho_0-\rho_1$, achieving the trace distance. Let us define what
we mean by an ancilla-free bit commitment:

\begin{defi}[Ancilla-Free One-Way Bit Commitment Protocol, AFBC]
A one-way bit commitment protocol in which we restrict the cheating strategies to ones
which have no access to additional quantum systems, i.e. ancillas. Thus cheating strategies consist
of local unitary transformations and complete von Neumann measurements.
\end{defi}

\section{One-way Bit Commitment with Superselection Rules}

To proceed with this analysis, we must stipulate how the
one-way bit commitment protocol is constrained by superselection rules \cite{AS:ssrule}.
\footnote{The distinction
between selection rules and superselection rules is not always very clear. At an informal level one may think
that superselection rules are due to fundamental laws in nature and thus never to be violated,
whereas selection rules are of a more relative nature, valid for the particular energy scales at hand, or due
to technological/practical constraints. A different and more precise definition was given in the original paper \cite{WWW:ss}. There superselection
rules are defined as selection rules, i.e. a dynamical conservation laws, with the additional restriction that
off-diagonal matrix elements in the conserved quantum number basis cannot be distinguished by measurement.
These definitions of selection and superselection rules seem to leave open the possibility for an
initial condition that is a superposition of different quantum numbers. In this section we specify
in detail what we mean by bit commitment constrained by the superselection rules. }
We also refer to \cite{bw:selection} for a detailed description of how superselection rules and their associated symmetry groups impose constraints.

We assume that Alice and Bob's actions have to obey a local superselection rule which can
be given by a Hermitian operator $K_A$ and $K_B$. We will discuss two examples here: 1) $K_{A/B}$
is the local particle number $N_{A/B}$. 2) $K_{A/B}$ is a local total angular momentum operator $J^2_{A/B}$; this is a
`nonabelian' case (the symmetry group is that of space
rotations).

Here then are the restrictions that we will require for the
one-way bit commitment. These restrictions include the possible
cheating strategies of the parties:
\begin{enumerate}
\item The states in the token+proof Hilbert space in Alice's lab
are eigenvectors of $K_A$ with the same eigenvalue, say $k$. Thus,
$K_A\ket{\chi_0}=k\ket{\chi_0}$ and
$K_A\ket{\chi_1}=k\ket{\chi_1}$. If Alice cheats and she chooses
to create other states, then these should also be eigenstates of
$K_A$. \footnote{In fact the security of both our schemes does not
change if we permit Alice to create {\em any} state $\chi$.} \item
Any action that Alice performs before the commit phase must
involve unitary transformations that leave the eigenvalue $k$
unchanged, i.e., $[U,K_A]=0$. \item After the commit phase, any
unitaries performed by Alice must satisfy $[U,K_A]=0$. Likewise
for Bob when he is given the token: $[U,K_B]=0$. \item Any
measurements that Bob performs must respect the superselection
rule.  Thus, in the commit phase (when Bob might wish to cheat)
any measurement projector $\Pi$ of Bob on the token space must
satisfy $[\Pi,K_B]=0$.  In the unveiling phase, the same must be
true for the joint token+proof Hilbert space in Bob's possesion:
$[\Pi,K_B]=0$.
\end{enumerate}

If the cheating strategy is not required to be ancilla-free, Bob
could create a state of definite quantum number $k_B$ in his lab
before he receives the token. After he has received the token, he
should respect the conservation rule on his total token+ancilla
space, but the ancilla may still help him to do a better
measurement.


Let us use the following notation: since a one-way BC
protocol is completely specified once the orthogonal pair
$\ket{\chi_{0,1}}$ is agreed upon, we will refer to this protocol
as $\mbox{BC}(\ket{\chi_{0,1}})$.  If the protocol has the above
superselection rule restrictions imposed involving operator $K$, we
will refer to this protocol as $\mbox{BC}_K(\ket{\chi_{0,1}})$.
If the protocol is forced to be ancilla-free we write $\mbox{AFBC}_K(\ket{\chi_{0,1}})$.

\section{Our results}

We find that in the presence of particle number superselection rules, the security of one-way BC is rigorously
unchanged, see Section \ref{pn}:
\begin{eqnarray}
&&\forall\ket{\chi_{0,1}}\,\mbox{such
that}\,N\ket{\chi_{0,1}}=n\ket{\chi_{0,1}},\nonumber\\
&&G^{\mbox{max}}(\mbox{BC}_N(\ket{\chi_{0,1}}))=
G^{\mbox{max}}(\mbox{BC}(\ket{\chi_{0,1}}))=G^{\mbox{max}}(\mbox{AFBC}(\ket{\chi_{0,1}})),\nonumber\\
&&C^{\mbox{max}}(\mbox{BC}_N(\ket{\chi_{0,1}}))=
C^{\mbox{max}}(\mbox{BC}(\ket{\chi_{0,1}}))=C^{\mbox{max}}(\mbox{AFBC}(\ket{\chi_{0,1}})).\label{abel}
\end{eqnarray}

However, in the presence of angular momentum superselection rules
(operator $J^2$; a `nonabelian' case), a one-way BC with arbitrarily
high security can be devised {\em if we force the scheme to be ancilla-free}
\begin{eqnarray}
&&\exists\,\ket{\chi_{0,1}(j)},\,\,\mbox{such
that}\,J^2\ket{\chi_{0,1}(j)}=j(j+1)\ket{\chi_{0,1}},\nonumber\\
&&G^{\mbox{max}}(\mbox{AFBC}_{J^2}(\ket{\chi_{0,1}(j)}))=0,\nonumber\\
&&C^{\mbox{max}}(\mbox{AFBC}_{J^2}(\ket{\chi_{0,1}(j)}))\rightarrow 0 \mbox{ for $j \rightarrow \infty$.}\label{nabel}
\end{eqnarray}
We do not have a rigorous proof of the statement about the
asymptotic behavior of
$C^{\mbox{max}}(\mbox{AFBC}_{J^2}(\ket{\chi_{0,1}}))$, but we
conjecture a formula for the fidelity function which coincides
with numerical data up to $j=11$, and which implies the asymptotic
security just stated; see Section \ref{angmom}.

\section{Particle Conservation}
\label{pn}
We now prove Eq. (\ref{abel}). In Ref. \cite{mayers:ss} Mayers first proved the simplest case in which the fact that the
protocol is completely sealing makes it completely unbinding, i.e. Alice can always change her commitment. The arguments here are
 a straightforward extension of this simple case. The number
operator is additive over tensor product Hilbert spaces, so we can
write
\begin{equation}
N=N_t+N_p.
\end{equation}
This means that if, as we assume,
$N\ket{\chi_{0,1}}=n\ket{\chi_{0,1}}$, then these states must have
the form
\begin{equation}
\ket{\chi_b}=\sum_{i_t,i_p,m=0}^n
c_b(i_t,i_p,m)\ket{i_p,n-m}_p\ket{i_t,m}_t,\label{num1}
\end{equation}
that is, if the token system has $m$ particles, the proof system
must have $n-m$ particles.  The labels $i_p$ and $i_t$ denote other quantum numbers characterizing the
states. It is understood that the range of the
$i_t$ and $i_p$ sums can depend on $m$ (i.e., they depend on the
local particle number).

The security parameters defined above depend only on
the reduced density operators on the token subsystem of the two
states.  For the states of the form of Eq. (\ref{num1}), these can
be written as
\begin{equation}
\rho_b=\sum_{i_t,j_t,m=0}^n
\sum_{i_p}c_b(i_t,i_p,m)c_b^*(j_t,i_p,m)\ket{i_t,m}\bra{j_t,m}
=\bigoplus_{m=0}^n p_{b,m}{\hat\sigma}_{b,m},\label{osum}
\end{equation}
where $p_{b,m}$ is the probability of
each $m$, and ${\hat\sigma}_{b,m}$ is a normalized density
operator in each $m$ sector: $\tr\,{\hat\sigma}_{b,m}=1$.

Consider what happens if Bob cheats. The ideal optimal measurement
for Bob that he can do to achieve Eq. (\ref{dist}) is a complete von
Neumann measurement in the eigenbasis of $\rho_0-\rho_1$. But, because of
Eq. (\ref{osum}), this eigenbasis is also an eigenbasis
of the particle number of the token system and given the superselection rule,
Bob is allowed to do this optimal measurement. So, Bob can gain exactly the same amount of
information in the protocol constrained by superselection rules,
$G^{\mbox{max}}(\mbox{BC}_N(\ket{\chi_{0,1}}))=
G^{\mbox{max}}(\mbox{BC}(\ket{\chi_{0,1}}))$.

Next, we consider what happens if Alice cheats. We will show that \linebreak
$C^{\mbox{max}}(\mbox{BC}_N(\ket{\chi_{0,1}})=C^{\mbox{max}}(\mbox{BC}(\ket{\chi_{0,1}})$.
First of all, due to the block-diagonal character of $\rho_0$ and $\rho_1$
we can write
\begin{equation}
C^{\mbox{max}}(\mbox{BC}(\ket{\chi_{0,1}})={1\over
2}\tr\,|\sqrt{\rho_0}\sqrt{\rho_1}|
={1\over 2}\sum_{m=0}^n\sqrt{p_{0,m}p_{1,m}}F(\hat{\sigma}_{0,m},\hat{\sigma}_{1,m}).\label{uhl3}
\end{equation}
On the other hand, Uhlmann's theorem for the fidelity function $F$ is also
written as:
\begin{equation}
F(\rho_0,\rho_1)=\max_{U_p}|\bra{\chi_0}U_p\otimes
I_t\ket{\chi_1}|=\max_{U_p}\Re(\bra{\chi_0}U_p\otimes
I_t\ket{\chi_1}).\label{uhl}
\end{equation}
We arrive at this last form by recognizing that $U_p$ can have any
arbitrary global phase. In Ref. \cite{SR:bc} it is shown that $C=|\bra{\chi_0}U_p\otimes
I_t\ket{\chi_1}|/2$ can be achieved if Alice creates the
$b$-independent state
\begin{equation}
\ket{\chi} \propto (\ket{\chi_0}+e^{-i\,{\mbox{arg}}(\bra{\chi_0}U_p\otimes
I_t\ket{\chi_1})}\ket{\chi_1})
\label{cheatstate}
\end{equation}
prior to the commit phase.  If she decides during the commit phase that $b=0$,
she leaves that state unchanged and sends it to Bob; if she wants $b=1$, she applies
the optimal $U_p^\dagger$ of Eq. (\ref{uhl}) to the proof system that she still holds.

But in the presence of the charge superselection rule, it is not
possible to do the unconstrained maximization of Eq. (\ref{uhl});
we must respect constraints 2. and 3. above imposed by the
superselection rules, which require that $U_p$ be block diagonal in the
charge index. We observe that Alice can create the cheating state $\chi$ of
Eq. (\ref{cheatstate}) while respecting the superselection rule. Thus we can write $C^{\mbox{max}}$ in this case as a
constrained maximization:
\begin{equation}
C^{\mbox{max}}(\mbox{BC}_N(\ket{\chi_{0,1}}))={1\over
2}\max_{U_p=\bigoplus_{m=0}^nU_{m,p}}\Re(\bra{\chi_0}U_p\otimes
I_t\ket{\chi_1})\label{uhlcon}
\end{equation}

In order to evaluate this expression, we rewrite the states of Eq. (\ref{num1}) as
\begin{equation}
\ket{\chi_b}=\sum_{m=0}^n\sqrt{p_{b,m}}\ket{\chi_{b,m}}.\label{newdec}
\end{equation}
Putting (\ref{newdec}) into (\ref{uhlcon}) and working out the expression gives
\begin{eqnarray}
C^{\mbox{max}}(\mbox{BC}_N(\ket{\chi_{0,1}}))={1\over
2}\sum_{m=0}^n\sqrt{p_{0,m}p_{1,m}}F(\hat{\sigma}_{0,m},\hat{\sigma}_{1,m}).\label{uhl2}
\end{eqnarray}
In other words, we obtain the claimed equality
$C^{\mbox{max}}(\mbox{BC}_N(\ket{\chi_{0,1}})
=C^{\mbox{max}}(\mbox{BC}(\ket{\chi_{0,1}})$.

We will now find a very different situation for angular-momentum conservation.
The idea is to find two states $\chi_{b}$ for $b=0,1$ such that the local density
matrices for Bob, $\rho_b$, look the same given the superselection rule. This implies that the diagonal elements
of $\rho_b$ in the local definite angular momentum basis must be the same. The off-diagonal elements can be
different however, and we can try to adjust these free parameters so as to limit Alice's cheating strategies.
In the case of particle conservation there is no room to adjust free parameters since the off-diagonal elements
of $\rho_b$ in the number basis are always zero, see Eq. (\ref{osum}). This is the essential difference between the abelian and
the non-abelian case.

\section{Angular Momentum Conservation}
\label{angmom} We introduce a family of one-way bit commitment
protocols, one for each total angular momentum quantum number $j$
($2j \in \Bbb{Z}^+$), although we will only discuss the integer
case in detail). Alice's two states $\chi_{0,1}$ are states with
total angular momentum $j_{tot}=j$ and $m_{tot}=j$. We denote the
Clebsch-Gordan coefficients as
$C(j_A,m_A,j_B,m_B,j_{tot},m_{tot})$ where $j_A,m_A$ are the
quantum numbers of the `proof' spin (kept by Alice, initially) and
$j_B,m_B$ are the quantum numbers of the `token' spin (sent to
Bob). Unlike the number operator, total angular momentum is not an
additive quantity (herein lies the crucial difference between the
security of the two cases), i.e. the two local angular momenta
$j_A$ and $j_B$ can give rise to total angular momentum between
$|j_A-j_B|$ and $j_A+j_B$. We take states for fixed $j$ to be
\begin{equation}
\chi_0(j)=\sum_{j_B=0}^{2j}\sqrt{\beta_{j_B}}\sum_{m_A=-j}^j
C(j,m_A,j_B,j-m_A,j,j)\ket{j,m_A}_p\ket{j_B,j-m_A}_t,
\end{equation}
\begin{equation}
\chi_1(j)=\sum_{j_B=0}^{2j}(-1)^{j_B}\sqrt{\beta_{j_B}}\sum_{m_A=-j}^j
C(j,m_A,j_B,j-m_A,j,j)\ket{j,m_A}_p\ket{j_B,j-m_A}_t.
\end{equation}
Of course, the $z$-component of the local spin cannot be larger
than the total local spin; we impose this in the above equations
by taking the  the Clebsch-Gordan coefficients
$C(j,m_A,j_B,j-m_A,j,j)$ to be zero if $j-m_A > j_B$.  The
$\beta$-coefficients obey the following constraints:
\begin{eqnarray}
\sum_{j_B=0}^{2j}\beta_{j_B}=1, \\
\sum_{j_B=0}^{2j}(-1)^{j_B}\beta_{j_B}=0,\\
\forall\,j_B,\;\beta_{j_B} \geq 0.
\label{constrs}
\end{eqnarray}
The first equation enforces the normalization and the second
equation enforces the orthogonality of $\chi_0$ and $\chi_1$.
Before considering Alice's strategies, which will give us
additional equations for the $\beta$-coefficients, let us consider
Bob's cheating strategies. Due to the superselection rule his
measurement has to be diagonal in the total angular momentum of
his particle which implies that he will not be able to detect any
difference in the terms of $\rho_0$ and $\rho_1$ that are
off-diagonal in the $j_B$-basis. The states $\chi_b(j)$ are chosen
such that the on-diagonal terms of $\rho_0$ and $\rho_1$ are
identical. This implies that for all $j$,
$G^{\mbox{max}}(\mbox{AFBC}_{J^2}(\ket{\chi_{0,1}}))=0$.

In order to analyze Alice's cheating ability, we work out the expression for the fidelity $F$ in
Eq. (\ref{fid}) which provides an upper bound on her cheating strategies. Note that the expressions for
$\chi_b$ have the Schmidt form:
\begin{equation}
\ket{\chi_b}=\sum_{m_A}\sqrt{\lambda_{m_A}}
\ket{j,m_A}\ket{\phi_b^{m_A}},\label{ssm}
\end{equation}
where $\phi_b^{m_A}$ are normalized orthogonal vectors:
\begin{equation}
\ket{\phi_b^{m_A}}=\frac{1}{\sqrt{\lambda_{m_A}}}\sum_{j_B=0}^{2j}(-1)^{b.j_B}\sqrt{\beta_{j_B}}
C(j,m_A,j_B,j-m_A,j,j)\ket{j_B,j-m_A}.
\end{equation}
The Schmidt coefficients in Eq. (\ref{ssm}) are independent of the
bit $b$:
\begin{equation}
\lambda_{m_A}=\sum_{j_B} \beta_{j_B} C^2(j,m_A,j_B,j-m_A,j,j).
\end{equation}
Also note that for these states
\begin{equation}
\langle\phi_0^{m_A}\ket{\phi_1^{m'_A}}=\frac{1}{\lambda_{m_A}}K_{m_A}\delta_{m_A m'_A},
\end{equation}
where
\beq
K_{m_A}=\sum_{j_B=0}^{2j}(-1)^{j_B}\beta_{j_B}
C^2(j,m_A,j_B,j-m_A,j,j).
\eeq
With these tools, Uhlmann's fidelity can be written as
\beq
F(\rho_0,\rho_1)=\tr \sqrt{\sqrt{\rho_0}\rho_1\sqrt{\rho_0}}=\sum_{m_A} |K_{m_A}|=\sum_{m_A} s_{m_A} K_{m_A},
\eeq
where $s_{m_A}={\rm sign}(K_{m_A})$ and thus
\beq
F=\sum_{m_A=-j}^j s_{m_A} \sum_{j_B=0}^{2j}(-1)^{j_B}\beta_{j_B}
C^2(j,m_A,j_B,j-m_A,j,j).
\label{Ffinal}
\eeq
We arrive at a piecewise linear program: determine the vector $\vec{\beta}$ which minimizes
$F$ under the constraints given by Eq. (\ref{constrs}).
For completeness, let us state the Clebsch-Gordan coefficients as they appear in the expression for $K_{m_A}$:
\begin{eqnarray}
C^2(j,m_A,j_B,j-m_A,j,j)&=&{{-m_A+j+j_B\choose
-m_A+j}{m_A+j\choose m_A-j+j_B}\over{2j+j_B+1\choose
2j+1}},\,\,\,0\leq j_B\leq 2j,\,1\leq m_A\leq
j,\,m_A\geq j-j_B,\nonumber\\
C^2(j,m_A,j_B,j-m_A,j,j)&=&0,\,\,\,\,\,\,\,\,\,0\leq j_B\leq
2j,\,1\leq m_A\leq j,\,m_A< j-j_B.
\end{eqnarray}
(That is, a binomial coeficient outside its usual range should be
taken to be zero.)

\subsection{Numerical Analysis}

For each integer $j$ we have investigated this piecewise linear
program numerically. (We do not report the half-integer $j$
results here, they work out similarly.). We have done the minimization up to $j=11$ and we
observe the following patterns. As it turns out, in all solutions we find that
$\beta_{k}=0$ for $j+2 \leq k \leq 2j$. Secondly, in every case, the terms in Eq. (\ref{Ffinal})
for $m_A > 0$ are identically zero.  This gives us a set of linear
equalities for $\beta_{j_B}$:
\beq
\forall\,m_A:\; 1\leq
m_A\leq j,\;\sum_{j_B=0}^{2j}(-1)^{j_B}\beta_{j_B} C^2(j,m_A,j_B,j-m_A,j,j)=0.
\eeq
Thirdly, in alternating cases (j even or j odd), we find that
$s_{-1}$ and $s_0$ are $\{+1,-1\}$ and $\{-1,+1\}$.  In
addition, the remaining terms in Eq. (\ref{Ffinal}), for $m_A<-1$,
are also identically zero. For example, the solutions
for $j=1,2,3$ are
\begin{itemize}
\item $j=1$: $\beta_{0,1,2}=\{ {2\over 9},{1\over 2},{5\over
18}\}$, leading to the sign assignments $s_{-1,0,1}=\{+1,-1,x\}$
where $x$ implies that $s_{1}$ is irrelevant in the minimization. This leads to $F={1\over 3}$.
\item $j=2$: $\beta_{0,1,2,3,4}=\{ {3\over 20},{9\over
25},{7\over 20},{7\over 50},0\}$, leading to the sign assignments
$s_{-2,-1,0,1,2}=\{x,-1,+1,x,x\}$. For these
$\beta$ values, $F={1\over 10}$.
\item $j=3$: $\beta_{0,1,2,3,4,5,6}=\{ {4\over 25},{2\over
7},{78\over 245},{3\over 14},{33\over 490},0,0\}$, leading to the
sign assignments \linebreak $s_{-3,-2,-1,0,1,2,3}=\{x,x,+1,-1,,x,x,x\}$.
For these $\beta$ values, $F={1\over 35}$.
\end{itemize}

Most importantly, we have observed that for all cases up to $j=11$, we find optimal
values of $F$ that agree with the simple formula:
\begin{equation}
F=\left [{2j+1\choose j+1} \right]^{-1}.
\end{equation}
As is clear from this formula, which we conjecture to correspond
to a feasible solution for the $\vec{\beta}$-vector for all $j$,
$F$ goes to zero exponentially fast as $j\rightarrow\infty$ which
implies the security of the protocol.

\section{Acknowledgements}


We are grateful for the support of the National Security Agency
and the Advanced Research and Development Activity through
contract DAAD19-01-C-0056.

\bibliographystyle{hunsrt}
\bibliography{refs}

\end{document}